\documentstyle[preprint,aps]{revtex}
\begin{document}
\draft
\title{Evolution of wave packets in quasi-1D and 1D random media: diffusion versus
localization}
\author{F.M. Izrailev$^{1,2}$ \thanks{
email addresses: izrailev@ inp.nsk.su; izrailev@physics.spa.umn.edu},
Tsampikos Kottos$^2$, A. Politi$^3$, and G.P. Tsironis$^2$}
\address{$^1$ Budker Institute of Nuclear Physics,\\
630090 Novosibirsk, Russia}
\address{$^2$ Department of Physics, University of Crete and Research Center \\
of Crete, P.O. Box 2208, 71003 Heraklion-Crete, Greece}
\address{$^3$ Istituto Nazionale di Ottica 50125 Firenze and \\
INFN-Firenze, Italy}
\date{\today }
\maketitle

\begin{abstract}
We study numerically the evolution of wavepackets in quasi one-dimensional
random systems described by a tight-binding Hamiltonian with long-range
random interactions. Results are presented for the scaling properties of the
width of packets in three time regimes: ballistic, diffusive and localized.
Particular attention is given to the fluctuations of packet widths in both
the diffusive and localized regime. Scaling properties of the steady-state
distribution are also analyzed and compared with a theoretical expression
borrowed from the one-dimensional Anderson theory. Analogies and differences
with the kicked rotator model and the one-dimensional localization are
discussed.
\end{abstract}

\vspace{0.4cm}

\section{\bf Introduction}

The main approach to a statistical description of the spectra in complex
quantum systems originates from the pioneering work of Wigner \cite{P65},
who conjectured that random matrices could represent the simplest meaningful
model for studying heavy nuclei. Currently, random matrix theory (RMT) has
become a very effective tool in a large variety of physical applications.
Until recently the matrices in this theory were assumed to be homogeneous,
i.e. all matrix elements were taken to have identical statistics. This
simplification is mainly dictated by mathematical reasons since the
corresponding ensembles of random matrices are rotational invariant, a
property that simplifies the theoretical analysis.

For a long time, the RMT had no concrete physical basis, in the sense that
conditions for its applicability were not specified. It was believed that
the systems under consideration had to be extremely complex in order to have
a good agreement with the predictions of the RMT \cite{P65,M67,Brody}. The
situation has changed with the progress of the so-called quantum chaos
theory which deals with dynamical Hamiltonian systems exhibiting chaotic
motion in the classical limit. One of the main results of this theory is
that in the extreme case, when classical motion is strongly chaotic and no
influence of quantum localization is taken into account, statistical
properties of both spectra and eigenfunctions are well described by the RMT.
This statement has been thoroughly studied and confirmed both for autonomous
systems like chaotic billiards \cite{BG84} and for time-dependent models
like the Kicked Rotator model (KR) \cite{CCFI79,I90} and the Kicked Tops 
\cite{H91}. Moreover, there are many physical examples where ``complexity''
of a quantum system is not maximal, but nevertheless a statistical
description applies pretty well.

To describe the consequences of quantum localization in the presence of
strong classical chaos, a new type of random matrices has been introduced
when studying the KR \cite{CGIS90}. The distinctive feature of these
matrices is their band-like structure, which is related to the finite range
of interactions in a given basis. In a sense, the ensembles of Band Random
Matrices (BRM) can be regarded as an extension of the conventional random
ensembles, since the latter are recovered by just setting the band size $b$
equal to the matrix size $N$. Currently, the interest for BRM raised
significantly due to their close relationship with quasi-1D models with
random potentials \cite{FM94}, or 1D systems with long-range hoppings
between neighboring sites. In the 1D interpretation, the band size $b$
corresponds to the hopping range while in the quasi-1D case, it is
associated with the number of transverse channels for electron wave
propagation along a thin wire. Recent numerical and analytical studies of
BRM (see \cite{FM94,I95} and references therein), led to numerous results
regarding the structure of eigenstates. In particular, the localization
length has been shown to depend only on the scaling parameter $b^2/N$, so as
the statistical properties of the eigenstates that are directly related to
the fluctuation properties of the conductance.

However, much less is known about the evolution of wave packets in models
described by BRM. One should note that even in the ``simple'' case of
Anderson-type models in 1D only the short and the long time scales,
corresponding to ballistic spread and saturation of the packet width,
respectively, have been successfully studied \cite{NPF87}. In quasi-1D (or
1D with long-range hoppings) models, the picture is both more complicated
and more interesting with respect to the Anderson case. Indeed, while the
diffusion time scale is absent in 1D models of Anderson type, since the mean
free path is of the order of the localization length, classical diffusion
alters the ballistic spread in quasi 1D systems, before being eventually
suppressed by localization effects.

In dynamical systems, the influence of strong localization on classical
diffusion in momentum (or energy) space has been studied in detail in the
framework of the KR. It was found that the time scale of the wave-packet
spreading that is analogous to classical diffusion is much longer than the
logarithmic time scale over which the complete correspondence between
classical and quantum description holds \cite{CCFI79,I90,CIS81,CIS88}. The
entire diffusive process, including the final saturation proved to exhibit
remarkable scaling properties \cite{I90,CIS88}. In particular, the diffusive
time scale is proportional to the localization length of those eigenstates
which are involved in the dynamics. In view of the analogies existing
between dynamical and Anderson-type localization, the results obtained in
the study of wave packet evolution in the KR will represent the touchstone
for the present investigation of packet dynamics in quasi-1D disordered
models. However, because of the existence of basic differences as well (see
Refs. \cite{I90,CIS88,CC94,CC95,D96}), it is not presently clear to what
extent the similarity between dynamical and disordered systems can be pushed
forward.

In this paper we extend a previous study \cite{we} of initially $\delta$%
-like wave packets in a quasi-1D geometry, with particular attention to
their width and fluctuations on different time scales. We hope that our
detailed numerical investigations, providing direct information on the
dynamical properties of strong localization, will furnish also some insight
for future analytical studies.

The paper is structured in the following way. In Sec.~2, the model is
introduced and discussed together with the main properties of spectra and
eigenstates of BRM. In Sec.~3 both ballistic and diffusive time scales are
analyzed and the scaling properties of packets in terms of the band-width $b$
are established. The effect of noise in the destruction of coherence is also
briefly discussed. Section 3 terminates with the results for the suppression
of classical diffusion due to the localization of eigenstates. In section 4,
we focus on the problem of fluctuations of the shape of packets both for the
diffusive and relaxation time scales. In section 5, a phenomenological
description of the asymptotic shape of the packets is given, based on
results for the 1D Anderson model. Moreover, fluctuations along the
asymptotic profile of packets are studied. The conclusions are summarized in
Sec.~6.

\section{\bf Band Random Matrices: main properties}

\subsection{\bf Definition and applications}

Since its birth, random matrix theory has been mainly dealing with
statistical properties of ``full'' random matrices, for which all
off-diagonal matrix elements are independent and distributed according to
the same law. In physical applications this implies that interactions are
assumed to be so strong and complex that no other parameter, apart from the
symmetry of matrices, is to be taken into account. As a result, such
matrices are associated with the extreme case of maximal chaos which is
known to appear in various physical systems such as heavy nuclei, atoms,
metallic clusters, etc. Furthermore, full random matrices represent a good
model for the description of local statistical properties of spectra and
eigenstates in some range of the energy spectrum, typically, in the
semiclassical region.

On the other hand, the conventional RMT is both unable to describe important
phenomena such as localization of eigenstates, and to characterize the
spectra of physical systems influenced by strong localization effects. For
this reason, much attention has been recently paid to the so-called Band
Random Matrices which are characterized by the free parameter $b$ defining
the effective band width of a Hamiltonian. Such random matrices with
elements decaying away from the main diagonal, appear to provide more
realistic models for the Hamiltonian of ``complex'' quantum systems (see,
e.g., \cite{FM94,I95} and references therein). The simplest type of BRM is
given by matrices $H_{nm}$ with zero elements outside the band ($|n-m|>b)$,
while inside the band ($|n-m|\leq b$), matrix elements are assumed to be
independent and distributed according to a Gaussian law, 
\begin{equation}  \label{BRM}
{\cal P}(H_{nm})=\frac 1{\sigma _{nm}\sqrt{2\pi }}\exp(-H_{nm}^2/2\sigma
_{nm}^2),\,\,\,\,\,\,\,\,\,\sigma _{nm}^2 \equiv <H^2_{nm}> = \frac{\sigma
_0^2}2(1+\delta _{nm})\,\,
\end{equation}
where $\delta _{nm}$ is the Kronecker symbol, and $\sigma _0^2 = 2$ implying
that the variance of the off-diagonal matrix elements is equal to 1. The BRM
ensemble can be regarded as a generalization of the standard Gaussian
orthogonal ensemble (GOE) as the former reduces to the latter when $b=N$.
Analogous generalizations can be introduced for the unitary and symplectic
ensembles of band random matrices \cite{FM94}.

The limit case $b=0$ corresponds to diagonal matrices, while $b=1$
corresponds to tridiagonal Hamiltonians with both diagonal and off-diagonal
disorder. The latter case is well known in the physics of disordered media;
the main properties of such matrices are relatively well understood. The
general case of BRM where the variance of the matrix elements decreases with
the distance from the main diagonal, introduced in \cite{FM91}, is also
amenable to an analytic treatment. They are not, strictly speaking band
matrices, but an effective band size $b$ can be defined from the shape of
the envelope\cite{note}.

In what follows, we consider large values of $b\gg 1$ and assume that $N\gg
b^2$. The first condition implies a large number of nearby states coupled by
the interaction. The second condition allows to neglect finite size
corrections arising from the finiteness of the samples.

Considerable interest towards the ensemble of BRM was stimulated by the
investigation of the quantum behavior of periodically driven Hamiltonian
systems. A paradigmatic system in this class is the so-called Kicked Rotator
model. Indeed, the unitary matrix $U$ yielding the time evolution between
two consecutive kicks has, in the angular momentum representation, a
band-like structure with an effective band-size approximately equal to the
strength $k$ of the kicks. Outside the band, the matrix elements of $U$
decrease extremely fast, while inside the band they can be treated as
pseudo-random entries if the corresponding classical evolution is chaotic%
\cite{I90}. As a consequence, both spectra and eigenstates of BRM of the
type (\ref{BRM}) are expected to have statistical properties similar to
those of the KR. A number of data substantiate this belief \cite{FM94,I95}.

BRM play also an important role in the understanding of quasi-1D disordered
media. The connection has been established through the nonlinear sigma model
which, on the one hand, is known to provide an excellent description of
quasi-1D systems and, on the other, has been rigorously shown to be
reducible to the BRM model \cite{FM91}. Such relationships gave a boost to
the investigations of statistical properties of eigenstates and eigenvalues
of BRM as they also allow a better understanding of the properties of the
KR, as well as of quasi-1D and 1D models with long-range random interactions.

\subsection{\bf Density of States and Structure of Eigenstates}

As was first numerically shown in \cite{CMI90} and later analytically proved
in \cite{FM91}, the density of states $\rho (E)$ for infinite BRM ($N
\rightarrow \infty $) obeys the semicircle law 
\begin{equation}  \label{DOS}
\rho (E)=\frac 1{4\pi bv^2}\sqrt{8bv^2-E^2}\,\,;\,\,\,\,\,\,\,\,\left|
E\right| \leq R_0=v\sqrt{8b}
\end{equation}
with $\rho (E)=0$ for $\left| E\right| >R_0$ . The parameter $v$ is just the
standard deviation of the distribution of off-diagonal elements, $%
v^2=\left\langle H_{nm}^2\right\rangle =\sigma _0^2/2$; it does not
influence the statistical properties of the spectra as well as the structure
of the eigenstates, since it can be scaled out. For $b=N$, expression (\ref
{DOS}) reduces to the well known Wigner semicircle law derived in
conventional RMT.

In infinite BRM, all eigenstates $\varphi _E(n)$ are known to be
exponentially localized \cite{FM94,I95,CMI90}, 
\begin{equation}
\left| \varphi _E(n)\right| \sim \exp \left( -\frac{\left| n-n_0\right| }{
l_\infty (E)}\right) ;\,\,\,\,\,n\rightarrow \pm \infty  \label{EF}
\end{equation}
where $n_0$ is the ``center'' of an eigenfunction in the basis in which the
random matrices have been defined. The quantity $l_\infty (E)\,$ is the
localization length defined as the inverse of the asymptotic spatial
decay-rate of the amplitude of the corresponding eigenfunction. Numerically, 
$l_\infty (E)$ can be determined by implementing the transfer matrix method.
It is important to recall that the localization length $l_\infty$ describes
the decay of the eigenfunction only in the tail and not in the central
region of size $\approx b^2$. This region is characterized by an effective
number $l\sim l_\infty $ of ``principal components'' which are usually
defined in terms of the inverse participation ratio and of the so-called
entropic localization length $l_H$ (see \cite{I90,FM94} for details).

In finite samples, one more parameter comes into play, the rank $N$ of the
matrices. In such a case, all relevant properties of spectra and eigenstates
are parametrized by the ratio $\lambda =b^2/N$ \cite{FM94,I95}. Upon
changing $\lambda $, one can accurately follow the transition from the
completely localized $(\lambda \ll 1)$ to the delocalized $(\lambda \gg 1)$
regime. Numerous studies of BRM allowed to unravel the dependence of the
statistical properties of eigenfunctions on this scaling parameter. In
particular, in finite bases of size $N$, all eigenstates are extended if $%
\lambda \gg 1$, and all their properties are very similar to those for the
standard RMT. Our interest here, is devoted to the opposite limit $\lambda
\ll 1$ of very localized eigenstates where finite-size effects can be
neglected. Based on the results obtained for the KR \cite{I90,CMI90}, it was
predicted that the localization length $l_\infty $ is proportional to $b^2$.
A rigorous analysis \cite{FM91} has confirmed this prediction and
established the dependence $l_\infty \sim \rho ^2b^2$ of the localization
length on the energy. In the case of BRM with a general envelope function $%
a(k)$ for matrix elements $H_{nm}$, 
\begin{equation}
l_\infty (E)=2[1-({\frac E{R_0}})^2]B\,\,;\,\,\,\,\,\,\,\,\,B={\frac{%
\sum\limits_{k=-\infty }^\infty a(k)\,k^2}{[\sum\limits_{k=-\infty }^\infty
a(k)]^2}}\,\,\,;\,\,\,\,\,k=n-m  \label{locle}
\end{equation}
where $B$ is the second moment of the function $a(k)$. In case (\ref{BRM}),
i.e. for a sharp band of size $b$, one obtains $B=b^2/3$ and thus 
\begin{equation}
l_\infty (E)=\frac 23[1-({\frac E{R_0}})^2]b^2\quad .  \label{ll}
\end{equation}
At variance with the KR, the localization length of the eigenstates of BRM
depends on the energy $E$.

\subsection{\bf Numerical procedure}

Although considerable progress has been made in the description of the
eigenstate structure (see \cite{FM94,I95} and references therein), the
evolution of wave packets is still poorly understood even in the limit of
infinite samples. The mathematical model we consider below is the
time-dependent Schr\"{o}dinger equation on a 1D lattice, 
\begin{equation}
i\frac{dc_n(t)}{dt}=\sum_{m=n-b}^{n+b}H_{nm}c_m(t)\;,  \label{TDE}
\end{equation}
where $c_n(t)$ is the probability amplitude for an electron to be at site $n$
and $H_{nm}$ is a symmetric BRM of the type (\ref{BRM}). Equation (\ref{TDE}%
) has been integrated numerically using a finite-time-step ($dt\simeq
10^{-3}-10^{-4}$) fourth order Runge-Kutta algorithm on a self-expanding
lattice in order to eliminate finite-size effects \cite{MT94}. Whenever the
probability of finding the particle at the edges of the chain exceeded $%
10^{-15}$, $10b$ new sites were added to each edge. The initial condition
was taken to be a $\delta $-like state located in the middle of the chain,
i.e. $c_n(t=0)=\delta _{n,0}$. At each time step, the normalization
condition for the total probability, $\sum_n|c_n(t)|^2=1$, was checked
observing fluctuations smaller than $10^{-4}$.

A further check of the accuracy of our calculations has been performed by
reversing the time-axis direction after 2000 time units (for $b=10$). The
difference between the initial probability distribution and that obtained
after integrating for 2000+2000 units was found to be less than $10^{-13}$.
In all the cases discussed below, a large number of disorder realizations
has been considered (more than 150) in order to get rid of sample-to-sample
fluctuations.

\section{\bf Diffusion of wave packets}

The time-evolution of a quantum wave packet in the lattice is naturally
described by the mean square displacement, 
\begin{equation}  \label{MSD}
M(b,t)=\left\langle u(t)\right\rangle \equiv \left\langle
\sum_mm^2|c_m(t)|^2\right\rangle \,\, ,
\end{equation}
where $\left\langle ...\right\rangle $ stands for the average over different
realizations of the frozen disorder $H_{nm}$. The time dependence of $M(b,t)$
provides a qualitative description of the dynamical regime: a power-law
evolution, $M(t) \sim t^{\nu}$, with $\nu <1$ corresponds to a sub-diffusive
behaviour (hinting at a possible, eventual localization), $\nu=1$
corresponds to ordinary diffusion, while $\nu >1$ to super-diffusion ($\nu=2$
characterizes ballistic motion).

As mentioned above, BRM can be regarded as a good model for dynamical
quantum systems such as the KR in the region of strong classical chaos. In
the classical limit, this model exhibits an unbounded diffusion in angular
momentum space if the kick strength exceeds some critical value. It was
discovered that even in the deep semi-classical domain, quantum effects can
suppress classical diffusion \cite{CCFI79} giving rise to a phenomenon that
is closely related to Anderson localization of a quantum particle in random
potentials \cite{A58,MT61,B73}. This effect of ``dynamical localization''
was claimed to be experimentally observed in the ionization of hydrogen
subject to a monochromatic field \cite{CCGS87}. A formal connection with a
1D tight-binding model has been found in \cite{FGP82}, thus reviving a
general interest for localization in one-dimensional systems.

We have investigated the behaviour of $M(b,t)$ by numerically integrating
Eq.~(\ref{TDE}) for different values of $b$. The results of our analysis
have been compared with both theoretical predictions of the 1D Anderson
model and numerical findings for the KR.

\subsection{\bf Ballistic time scale}

The essential difference between periodic and disordered quantum lattice
structures mostly lies in the localization properties of their electronic
states. In the periodic case, all the states are perfectly extended Bloch
waves, while in strongly disordered samples, the states are asymptotically
localized in time because of quantum interference effects. However, even in
the latter case, there exists a ballistic regime, occurring on time scales
of the order of the elastic scattering time $t_b$, i.e the time for an
electron to move by an amount corresponding to the mean free path $l_m$. In
quasi-1D systems, $l_m$ is known to be equal to the number of transverse
channels.

In order to investigate the scaling behavior of the packet size with $b$, we
have numerically integrated Eq.~(\ref{TDE}) for very short times $\sim t_b$
and different values of $b$ in the range $b=20\div 45$. The behaviour of the
mean-square displacement $M(b,t)$ is reported in Fig.~1 with the scaling
assumption 
\begin{equation}
M(b,t)=b^2\tilde{M}(t\sqrt{b})\,\,.  \label{BTS}
\end{equation}
The very good data collapse confirms the scaling Anzatz. Eq.~(\ref{BTS}) can
be understood by estimating the ballistic time scale $t_b$. Let us start by
noticing that the leading contribution to the wavepacket spreading over
short time scales comes from the $2b$ sites which are directly coupled with
the site $n=0$ where the wave packet is concentrated at time $t=0$. By
evaluating the Schr\"{o}dinger equation and thereby determining $M(b,t)$,
one obtains 
\begin{equation}
M(b,t)=\sum_{n=-b}^bn^2|H_{n,0}|^2t^2\approx b^3t^2\;.  \label{bal4}
\end{equation}
The above type of evolution terminates when the average width $\sqrt{M}$ of
the packet becomes of the same order as the band-size $b$, so that farther
sites come into play. By substituting back in Eq.~(\ref{bal4}), one finds
that the ballistic spread occurs on the time scale 
\begin{equation}
t\,\leq \,t_b\,\approx \,{\frac 1{\sqrt{b}}}\,.  \label{tbscale}
\end{equation}
Accordingly, the ballistic time scale shrinks to zero for increasing the
interaction range $b$. Notice that the ballistic regime is entirely new with
respect to the analogous problem in the KR where, at small times, an
exponentially fast spread of the packet takes place.

\subsection{\bf Diffusive time scale}

In 1D Anderson-type models, wave packet saturation starts immediately beyond
the ballistic time scale, since the mean free path $l_m$ and the
localization length $l_\infty $ are of the same order ($l_\infty \approx
2l_m $). Accordingly, no intermediate diffusive regime can be observed. In
the case of large band-size, $b\gg 1$, the localization length $l_\infty
\sim b^2 $ is much larger than the mean free path $l_m\sim b$, so that a
diffusive time scale $t_D$ appears. In order to estimate $t_D$, we shall
follow the scaling arguments developed from the theory of quantum chaos
where they have been successfully introduced to explain the well known
``quantum suppression of classical diffusion'' in the KR \cite{CIS81}.

The first crucial observation is that the eigenfunctions $\varphi _E(n)$ are
exponentially localized in the standard basis for all energy values $E$ in
the spectrum (consequently, there is a pure-point spectrum $\rho (E)$). Let
us proceed by noticing that an initial state $c_n(t=0)$ excites an
effective, finite number $N_{eff}$ of eigenstates with corresponding
energies $E_i$. Accordingly, the spectrum of those eigenstates participating
to the evolution of the packet is characterized by a mean level spacing $%
\Delta _{eff}\sim R_0/N_{eff}$ where $R_0^2=8bv^2$ is the radius of the
semicircle, see Eq.(\ref{DOS}).

Therefore, for times $t<1/\Delta _{eff}$, the packet evolution does not
``feel'' the discreteness of the spectrum, which is resolved over longer
time scales and eventually leads to localization. A typical evolution is
reported in Fig.~2, where the dependence of the mean square displacement $%
M(b,t)$ for $b=12$ is shown on a very large time scale. However, as long as $%
t_b\leq t\leq t_D$, with 
\begin{equation}
\label{DTS}
t_D\approx N_{eff}/R_0\,\,  
\end{equation}
the motion of the particle is analogous to the standard (classical)
diffusion. This means that $M\approx Dt$, where $D$ is the diffusion
constant and $M$ can be interpreted as the square of the number of
effectively excited, unperturbed states. This regime is clearly seen in the
inset of Fig.~2, where the evolution of $M$ is reported in a doubly
logarithmic plot.

The quantity $M$ reaches its maximal value $M_{max}$ at $t\approx t_D$. The
value $M_{\max }^{1/2}$ is of the same order as the total number $N_{eff}$
of eigenstates that participate to the packet evolution. Let us finally
notice that the packet-width is asymptotically of the order of the
localization length of the eigenfunctions, i.e. $N_{eff} \sim l_{\infty}
\sim b^2$ (the energy dependence is here irrelevant and can be dropped).
Accordingly, the following scaling relations hold 
\begin{equation}  
\label{tdiff}
t_D \sim l_{\infty} / R_0 \; \quad \quad \quad D \sim l_{\infty} R_0 \quad ,
\end{equation}
where $R_0$ is the width of the spectrum. The second estimate in (\ref{tdiff}%
) corresponds to the well known relation between the localization length and
the diffusion coefficient in the theory of disordered solids, $l_\infty
\approx \pi \rho D $ where $\rho$ is the density of states (see, for
example, \cite{FM94}).

Eq.~(\ref{tdiff}) seems to suggest that the proper scaling of the time axis
is $b^{3/2}$. However, as we shall see in the next section, there is
convincing evidence that best scaling Ansatz, in comparison with the direct
integration of the Schr\"{o}dinger equation (\ref{TDE}), is $t/b^2$. Whether
the discrepancy is to be attributed to some weakness of the above arguments
or to an improper choice of the time units in the numerical procedure it is
not clear (in fact, everything works perfectly as if we had to refer to
dimensionless time units in Eq.~(\ref{TDE}) normalized to $\sigma_0$ - see
Eq.~(\ref{BRM}) - rather than to the width of the spectrum $R_0$).

\subsection{\bf Diffusion suppression and scaling properties}

One of the most important peculiarities of the time evolution of wave
packets in 1D and quasi-1D random potentials is the saturation of the width $%
M$ for $t\rightarrow \infty$. In analogy with the evolution of wave packets
in the KR \cite{I90} and from the localization properties of the stationary
problem (the localization length grows as $b^2$), one expects that for $b\gg
1$ the limiting value $M_\infty (b)$ grows as $b^4$. In order to confirm
this prediction, we have performed detailed numerical experiments in the
range $b=4 \div 12$. The asymptotic value $M_\infty (b)$ has been accurately
determined by averaging $M(b,t)$ over a long time after an initial
transient. From our data we have found that the dependence of $M_\infty$ on $%
b$ is slightly slower than expected, $M_\infty \sim b^\alpha$ with $\alpha
\approx 3.87 \pm 0.02$. This anomalous behaviour is presumably to be
attributed to the presence of finite band-size corrections which are not
negligible in the range of $b$ values that has been numerically investigated
($b \le 12$). Our results are reported in Fig.~3, where $M(b,t)$ and $t$ are
divided by the asymptotic value $M_\infty$ and its square root,
respectively. While the scaling Ansatz for $M$ follows straightforwardly
from the detailed knowledge of the localization properties, the rescaling of
time axis is mainly justified a posteriori from the resulting good data
collapse that can be observed in Fig.~3.

Anyway, the nontrivial evolution during the late stages of the diffusive
regime confirmed by a direct investigation. Since generic properties of
eigenstates in quasi--1D models have been found to be similar to those of
strictly 1D disordered models \cite{FM94}, it is natural to expect that the
similarity extends to the dynamics of wave packets as well. However, even in
1D-geometry, the analytical treatment is very difficult. Analytical results
are available only in the two opposite regimes, $t\ll 1$ and \thinspace $%
t\gg t_D$. For example, the asymptotic dependence of the mean squared
displacement $\langle u(t) \rangle \equiv \left\langle x^2(t)\right\rangle$
of packets in the long-time limit is given by \cite{NPF87}, 
\begin{equation}
\left\langle u(t)\right\rangle \simeq a_0l_m^2\left( 1-\frac{\ln (t/2t_m)} {%
t/2t_m} \right) ;\,\,\,\,\,\,t\gg t_m  \label{prig}
\end{equation}
where $l_m$ and $t_m$ are the mean free path and the corresponding time
between consecutive back scattering processes. This estimate is based on the
expression for the quantum diffusion coefficient obtained in \cite{B73}. The
logarithmic singularity in (\ref{prig}) follows from resonant transitions
occurring between pairs of the so-called Mott states \cite{M70,MD71}. Such
states have a peculiar structure characterized by two humps lying at a
distance much larger than their effective width. Since Mott states appear in
pairs, the corresponding energies are very close to each other and this
results in a resonant tunneling over large distances. The influence of Mott
states on electronic properties of disordered 1D models has been studied in 
\cite{GDP83}, where the clustering of energy levels was discovered and
attributed to these states.

The effect of these states has been included in the study of the long-time
behavior of wave packets in the KR \cite{C91}, where an expression similar
to Eq.~(\ref{prig}) has been introduced to describe the evolution of the
mean square displacement $M(t)$ in the momentum representation, 
\begin{equation}
\tilde{R}\equiv \frac{dM}{dt}\sim \frac{D^2 \ln \big(t/(2D)\big)} {\big(%
t/(2D)\big)^2} \,\,;\,\,\,\,\,\,\,\,t\gg 2D \quad .  \label{cohen}
\end{equation}
where $D$ is the classical diffusion coefficient. Numerical data seems to
confirm the expectations (Eq.~(\ref{cohen})), although the presence of very
large fluctuations prevent to draw a convincing conclusion. Despite the
better statistics of our data, the presence of the logarithmic correction
cannot be definitely assessed in BRM too.

Another approach to the problem of quantum diffusion in the presence of
strong localization has been recently suggested in \cite{C91a} (see further
developments in \cite{CC94}): it is essentially based on a phenomenological
diffusion equation for the Green function, which takes into account backward
scattering. At large times, the relaxation rate is given by \cite{CC94} 
\begin{equation}
\tilde{R}\sim \frac{D^2\ln {}^{3/2}\big( t/(2D)\big)}{\big( t/{2D}\big)^2}%
\,\,;\,\,\,\,\,\,\,\,t\gg 2D\quad ,  \label{chir}
\end{equation}
which differs from Eq.~(\ref{cohen}) by a further logarithmic factor. At the
moment, all available numerical data for the KR do not allow to draw a final
conclusion in favor of either expression. In any case, let us again remark
that quantum localization in the KR is of dynamical nature (there is no
randomness in the model), so that it is not clear to what extent it is
similar to the localization of Anderson type.

Instead of focussing on the question of the exact asymptotic dependence ($%
t\rightarrow \infty$) of the mean square displacement $M(b,t)$, it is, for
the time being, more useful to limit ourselves to provide an effective
description of the wide time region that includes also the crossover from
classical diffusion to complete saturation. In the absence of any theory, we
make use of the phenomenological expression suggested in \cite{BI88} (see
also \cite{I90}) 
\begin{equation}  \label{fit}
M(b,t) = M_\infty(b) \left( 1-\frac 1{\left( 1+t/{t_D}\right) ^\beta }%
\right) \; ,
\end{equation}
where $M_\infty$, $t_D$ and $\beta$ are the three independent parameters to
be determined. The first one is obviously obtained from the asymptotic
evolution, while the short-time classical diffusion (here, we neglect the
ballistic time scale which is indeed negligible for large $b$) provides a
further constraint to be fulfilled. In fact, for $t\ll t_D$, Eq.~(\ref{fit})
reduces to 
\begin{equation}  \label{Mdiff}
M(b,t)= \frac{\beta M_\infty}{t_D}t \, = \, Dt \; ,
\end{equation}
which allows expressing $t_D$ in terms of the last unknown $\beta$ which can
be determined by fitting the global behaviour of $M(b,t)$.

The main idea behind the phenomenological expression (\ref{fit}) is the
repulsion of the energy levels participating to the evolution of the packet.
As it was previously discussed, the diffusion rate is proportional to the
mean spectral density and it remains unchanged for $t\leq t_D$, according to
the uncertainty principle. However, for $t\geq t_D$ it decreases, since the
only eigenstates that continue to contribute (``operative eigenstates'') are
those whose energy level-spacing $s$ satisfies the relation $s\leq t_D/t$.
The relative number of such spacings (hence, the relative diffusion rate) is
given by the spacing distribution $p(s)$ for the operative eigenstates, 
\begin{equation}  \label{PS}
\frac{D(t)}D\sim \int_0^sp(s^{\prime })ds^{\prime }\sim s^{\beta +1} \sim
\left( \frac{t_D}t\right) ^{\beta +1} \quad ,
\end{equation}
where $t\gg t_D$ is assumed. The above time dependence for large times is
the core of the phenomenological expression (\ref{fit}) (see details in \cite
{I90}).

According to the above relation, the parameter $\beta $ characterizes an
effective repulsion between those eigenstates which are excited by the
initial wave packet. It is clear that these eigenstates strongly overlap. As
a result, the value of $\beta $ can be expected to be quite close to $1$.
Although these arguments are no longer valid for very long times, when the
level clustering \cite{D96,DS91} due to the influence of Mott states becomes
important, Eq.~(\ref{fit}) can still provide a sufficiently good description
of the packet dynamics. Numerical experiments done for the KR \cite{I90,BI88}
yield quite a small value of $\beta$ ($\beta \approx 0.3$). This result,
which is somehow contradictory with other studies (see the discussion in 
\cite{C91a}), is probably due to the insufficiently long times considered in
the simulations.

Our detailed numerical experiments with BRM, performed on a much longer time
scale and with high statistics, reveal quite a good correspondence with the
scaling dependence (\ref{fit}) (see Fig.~3). The best fit of Eq.~(\ref{fit})
gives the following values: $\beta \approx 0.8\,$ and $t_D \simeq 2 b^2$.
Since the values of the band size $b$ are not very large, the limiting value 
$M_\infty $ has been purposely rescaled to the same level for the different $%
b$-values. By neglecting the residual weak deviations from a perfect
scaling, we obtain $M_\infty \approx 1.9 b^2$. The most important point of
the above analysis is that the value of the repulsion parameter $\beta $ is
quite close to $1$. This means that even for very large times close to the
relaxation, the approximate power dependence $1/t^{0.8}$ for the difference $%
\Delta M = M_{\infty} (b) - M(b,t)$ mimics the correct dependence $\Delta M
\sim \ln (t)/t$ (see Eq.~(\ref{prig})). As a result, one can treat the
scaling dependence (\ref{fit}) as a good description of both classical
diffusion and its suppression due to strong localization of eigenstates.

The asymptotic localization of the wave packet is entirely a consequence of
the frozen character of the disorder in the Hamiltonian. However, in
reality, physical systems are also subjected to time-dependent noise (this
is, for instance, the case of applications to nuclear physics \cite{aurel}).
In the KR, the influence of a time-dependent noise was discussed for the
first time in \cite{OAH84} (see also the detailed investigation in \cite
{DG90}), finding that if the strength of the noise exceeds some critical
value, then it destroys coherent effects of quantum localization and pure
classical diffusion is recovered.

The addition of noise to the BRM provides an alternative method to determine
the diffusion coefficient $D$ from the direct computation of the linear
growth of $M(b,t)$. It is first interesting to notice that the computation
of $D$ cannot be easily performed in practice without the presence of noise.
The reason is that in the absence of a time-dependent noise, corrections to
the linear behaviour of $M(b,t)$ arise already at short times, thus
preventing an accurate determination of the coefficient of the linear
growth. In other words, coherent effects of quantum localization come into
play even on the time scale of classical diffusion and this results in a
rather smooth transition from classical diffusion to complete relaxation
(see Fig.~3).

In order to estimate the critical value of the noise-strength $g^2_{cr}$
which destroys completely quantum coherence and leads to pure classical
diffusion, one needs to compare the shift of levels induced by the
additional noise, with the mean level spacing between operative eigenstates.
Since the latter turns out to be proportional to $1/b^2$, one can see that
if the shift $\Delta E\approx g^2t_D$ is larger than $1/b^2$, localization
will be completely destroyed. Accordingly, effects of quantum coherence
should be observable only when the condition $g^2\geq 1/b$ is satisfied. Our
numerical simulations confirm this estimate: the data in Fig.~4, which refer
to $b=8$, show that $D/b^2\approx 1.8$ in good agreement with the value
found from the fit of the scaling dependence given by (\ref{fit}). We would
like to stress that diffusion due to the noise occurs also for $g < g_{cr}$
however, the rate of such diffusion is different from that given by the
classical diffusion determined in the limit $b \rightarrow \infty$ (for
details see, e.g., \cite{DG90}).

\section{\bf Fluctuations of packet width}

In any statistical process, the analysis of average quantities provides a
limited description of the underlying properties. At least the variance
should be considered in order to achieve a more complete characterization of
the phenomenon of interest. In the present case, we shall consider the
fluctuations of $M(b,t)$ in both the diffusive ($t_b\leq t\leq t_D$) and the
relaxation ($t\gg t_D)$ regime. The relevant quantity to be determined is
the size of sample-to-sample fluctuations, 
\begin{equation}
\Delta M(b,t)\equiv \left( \left\langle u(t)^2\right\rangle
\,\,-\,\left\langle u(t)\right\rangle ^2\right) ^{1/2}  \label{DM}
\end{equation}
where the brackets $\left\langle ...\right\rangle$ denote the average over
different realizations of the Hamiltonian $H_{nm}$. A meaningful way to
present the numerical data is through the relative amplitude $\mu \equiv
\Delta M/M$ of the fluctuations. Its scaling properties with $b$ are
discussed in the two following subsections.

\subsection{Diffusive time scale}

As the diffusive time scale is relatively short, we have been able to
determine $\mu$ for quite large $b$-values (namely, $b \le 45$), so that
finite band-size corrections should be definitely negligible. In our
numerical experiments, we have integrated Eq.~(\ref{TDE}) up to a time $%
t_m=b^2/10$ for more than 1000 realizations of the disorder. The results are
reported in Fig.~5 under the scaling assumption 
\begin{equation}  \label{df}
\mu(b,t) \approx b^{-\eta} \overline \mu(t/b^2) \quad .
\end{equation}
By performing a least square fit of $\log \mu$ versus $\log b$ at fixed time 
$t = b^2/10$, well inside the diffusive regime, we have estimated $\eta$
which turns out to be approximately 0.7 (see the inset in Fig.~5). One
should notice the substantial disagreement with the value of the exponent
recently obtained in the KR \cite{CCI94}, $\eta \approx 1.0$. This scaling
parameter has been conjectured to be related to the mesoscopic fluctuations
of the diffusion coefficient \cite{CCI94}. However, the connection has not
been entirely clarified.

\subsection{\bf Relaxation time scale}

The next important issue concerns fluctuations around the so-called steady
state distribution of the wave packet in the asymptotic regime $t\gg t_D$.
If one assumes that the steady state distribution of $c_n(t \rightarrow
\infty )$ is characterized by an ergodic spread of the packet over some
finite size $N_s$, and if the components $c_n$ are statistically
independent, then 
\begin{equation}  \label{NS}
\mu \equiv \frac{\Delta M}M \approx \frac 1{\sqrt{N_s}}\,\,\,.
\end{equation}
Since the components $c_n$ are directly related to the amplitudes of
eigenstates and the latter are expected to be random on the scale of their
localization length, one can conclude that $N_s\approx \, \langle l_\infty
(b) \rangle \sim b^2$, i.e. $\mu \approx 1/b$. Surprisingly, our numerical
data in the range $b=4 \div 12$ indicate that the scaling has a definitely
different form, namely 
\begin{equation}  \label{delta}
\mu(t/b^2) \approx \frac 1{b^\delta }
\end{equation}
with $\delta \approx 0.7$ (see Fig.~6). A detailed study shows that when the
value of $\delta$ is varied by $\pm 0.1$, the superposition of the various
curves on the plateau gets appreciably worse. As an additional check, we
have performed a least square fitting of $\mu$ versus $b$ after averaging
the curves over times $t>20b^2$. The fit presented in the inset of Fig.~6,
confirms the value $\delta \approx 0.7\pm 0.01$ . One should note that this
value is in perfect agreement with the result found on the diffusive time
scale for the factor $\eta$, although there is no reason a priori to expect
such an equality over time scales where different physical mechanisms
control the packet evolution. We would like to stress that the anomalous
scaling described by Eq.~(\ref{delta}), is in close agreement with the
numerical data for the KR, where it has been observed that $\mu \sim
b^{-\delta}$ with an anomalous exponent $\delta \approx 0.6$ \cite{CC94}.
Note that in the KR, larger values of the effective parameter $b$ have been
reached.

In order to better disentangle the question of fluctuations, we have
investigated also the temporal behavior of $M(b,t)$. The main motivation for
this study is the comparison between sample-to-sample and temporal
fluctuations for a typical realization of the disorder. In practice, we have
integrated the Schr\"{o}dinger equation for a time up to $t=125b^2$,
discarding an initial transient time $t<$ $t_0=20b^2$ (which is sufficiently
long for $M(t)$ to saturate). The Fourier power spectrum $|U(\omega )|^2$ of 
$u(t)$ (let us recall that $M(t)= \langle u(t)\rangle$) signal has been then
averaged over more than 150 realizations. The results for $b=4,6$ and $8$
are reported in Fig.~7 in a doubly logarithmic plot. The best data collapse
is obtained by assuming that $\langle |U(\omega)|^2| \rangle \simeq
b^\varsigma $ with $\varsigma =6.6$. At ``high'' frequencies, $\langle
|U(\omega )|^2 \rangle $ exhibits a Lorentzian-type behaviour, which turns,
at low frequencies, into a weak divergence that reveals the presence of
non-trivial long-time correlations. In fact, by invoking the Wiener-Kintchin
theorem, the low-frequency tail in the spectrum of $\langle |U(\omega )|^2
\rangle$ can be connected with the relaxation properties of $M(t)$ towards
its asymptotic value $M_\infty$. More precisely, the power-law convergence
of the type $t^{-\beta}$ assumed in Eq.~(\ref{fit}) implies a power-law
divergence as $\omega^{-1+\beta}$ which is compatible with our data.
However, the low-frequency cut-off due to the finite time of our simulations
prevents drawing a definite statement about the presence of a truly
power-law divergence. Nevertheless, we can, at least, determine the
cross-over frequency $\omega _c$ separating the two temporal regimes, which
turns out to be $\omega _c\simeq 0.02/b^2$. Such a frequency is
approximately 100 times smaller than the mean spacing between the energy
levels of the eigenstates which effectively participate to the evolution of
the wave packet. This observation can be taken as an indirect confirmation
of the role played by Mott states in the long-time evolution. As it has been
already recalled, Mott states have quite a specific structure: they appear
in pairs and are characterized by two humps a distance $L$ apart. This leads
to a quasi-degeneracy of the order of $\Delta E \approx \exp(-L/l_\infty )$,
where $L$ is typically much larger than the localization length $l_\infty$.
Accordingly, over long time scales, a few Mott states may dominate the
packet dynamics.

For what concerns the scaling behavior of the spectrum $\langle |U(\omega
)|^2\rangle$ with respect to $b$, the estimated exponent $\varsigma \approx
6.6$ is in perfect agreement with (\ref{delta}). Indeed, the relation
between the spectral density (or power spectrum) and the variance of the
signal $M(t)$, implies 
\begin{equation}
(\Delta M(t))^2=\sum_\omega \left\langle |U(\omega )|^2\right\rangle
=\sum_{\omega b^2}b^{6.6}\left\langle |U(\omega b^2)|^2\right\rangle
=b^{6.6}(\Delta M(t/b^2))^2 \quad .  \label{power}
\end{equation}
By comparing Eq.~(\ref{power}) with Eq.~(\ref{delta}), one obtains again $%
\delta \approx 0.7$.

In Ref.~\cite{CC94} it was conjectured that the above anomalous scaling can
be considered as an indication of the fractal structure of the quantum
steady-state distribution $c_n(t\rightarrow \infty )$. More precisely, they
argued that the asymptotic shape can be described by an ensemble of only $%
N_s\sim l_\infty ^{0.6}$ statistically independent degrees of freedom ($N_s$
being the number of ``channels'' where the amplitude of the wavepacket is
essentially different from zero). The same conjecture can be raised in the
present case as well, although a direct check is an extremely hard task.

\section{\bf Steady-state distribution}

\subsection{\bf General discussion}

As mentioned above, the localization of all eigenfunctions (see Eq.~(\ref
{locle})) implies that for $t\gg t_D$ the quantum steady-state $f(n,t)\equiv
\left| c_n(t)\right| ^2$ fluctuates around an average profile $%
f_s(n)=\left\langle f(n,t)\right\rangle$. As the effective number of
eigenstates composing a single wavepacket is finite, the average profile
does depend on the disorder realization. However, in the limit $b \to \infty$%
, the number of statistically independent components diverges and
sample-to-sample fluctuations are expected to vanish. In that limit,
temporal and ensemble averages should coincide as long as the motion is
ergodic.

On the basis of diagrammatic techniques, many results have been obtained for
the steady state distribution $f_s(x)$ in continuous 1D models with white
noise potential (here, $x$ denotes the position of the electron). In
particular, an exponential decay $f_s(x)\sim \exp (-\left| x\right| /4l_m)$
has been predicted for the tails of $f_s(x)$ ($l_m$, being the mean free
path) \cite{B73}. A subsequent more accurate analysis \cite{GMR75} revealed
the presence of the prefactor $\left| x\right| ^{-3/2}$. Both findings are
included in the global expression derived in \cite{G76}, 
\begin{equation}  \label{gogolin}
f_s(x)=\frac{\pi ^2}{16l_m}\int_0^\infty \eta \,\hbox{sh} (\pi \eta ) \frac{%
(1+\eta ^2)^2}{(1+ch(\pi \eta ))^2} \exp\left( -\frac{1+\eta ^2}{4l_m}%
|x|)\right) d\eta \quad .
\end{equation}
In fact, the above expression implies that, close to the origin, the spatial
dependence is purely exponential, 
\begin{equation}  \label{orig}
f_s(x)\sim \exp(-|x|/l_m)\,;\,\,\,\,\,\,\,\,\,x\leq l_m
\end{equation}
while the asymptotic decay is described by 
\begin{equation}  \label{asymp}
f_s(x)\sim |x|^{-3/2} \exp(-|x|/4l_m)\,;\,\,\,\,x\gg 4l_m \quad .
\end{equation}
Therefore, the above two equations reveal that the decay rate $S(x) \equiv
(\ln \,f_s(x))^{\prime}$ (the prime denotes derivative with respect to the
argument), changes by a factor 4. One consequence of the non purely
exponential behavior is that the average size of the saturated packet is two
times smaller than the asymptotic localization length $\left\langle
|x|\right\rangle =2l_m$. It is interesting to notice also that the
asymptotic dependence (\ref{asymp}) is similar to that for the conductance
of 1D samples of finite size \cite{Z92} in the strong localized regime; in
this case $x$ is the ratio of the sample size with the localization length.
As no analytical results are available for quasi-1D systems, in the next
section we shall compare our numerical results with the above expressions,
by fitting the only free parameter $l_m$.

In any case, some information on the steady state distribution $f_s(n)$ can
be obtained from the structure of the eigenstates by exploiting the
following equality, 
\begin{equation}  \label{fs}
f_s(n)=\sum_m\left| \varphi _m(n_0)\varphi _m(n)\right| ^2 \quad ,
\end{equation}
where $\varphi _m(n)$ is the $n$-th component of the eigenstate with energy $%
E_m$ and $n_0$ is the position of the initial $\delta-$like packet.
Therefore, determining the asymptotic shape of a wave packet is tantamount
to determining the average correlation properties of single eigenstates.
Although no rigorous results are known in this direction, a phenomenological
approach allowed to shed some light on the closely related KR problem \cite
{I90,C91,BI88}. Because of the analogies between BRM and the KR, it is
instructive to compare the results for the steady distribution as well. A
rough estimation of the tails of $f_s(n)$ can be obtained in the following
way (see \cite{CIS88,CS86} for details). If one assumes the simple
exponential form $\varphi_m (n)\sim \exp (-|m- n|/l_\infty )$, for the $m$%
-th eigenstate, Eq.~(\ref{fs}) leads to the expression 
\begin{equation}  \label{chir1}
f_s(n)\sim \exp \left( -\frac{2|n-n_0|}{l_{\infty}} \right) \quad ,
\end{equation}
which, in turn, implies $l_s=l_\infty$, where $l_s$ is the localization
length of the asymptotic packet (defined from the probability amplitude,
i.e. taking the square root of Eq.~(\ref{chir1})). However, this result is
inconsistent with the numerical data for the KR which instead indicate that $%
l_s\approx 2l_\infty$ \cite{CS86}. To explain the latter result, it was
suggested to take into account the large fluctuations of the eigenstates
around their shape, 
\begin{equation}  \label{fluct}
\varphi_m(n) \sim \frac 1{\sqrt{l_\infty}}\exp \left( - \frac{|m-n|}{l_\infty%
}+\xi_{mn}\right) \quad ,
\end{equation}
where $\xi_{mn}$ is a Gaussian noise with the zero mean and a variance \cite
{CS86}, 
\begin{equation}  \label{locflu}
<(\Delta \xi _{mn})^2>\,=D_s|m-n|,\,\,\,\,\,\, D_s\sim 1/l_\infty \quad .
\end{equation}
By inserting the Ansatz (\ref{fluct}) in Eq.~(\ref{fs}) and averaging over
the noise term, it was found that 
\begin{equation}  \label{norm}
\left\langle \left| \varphi_m (n)\right| \right\rangle \, \sim \exp\left( - 
\frac{\left| n-m\right| }{2l_\infty }\right) \quad ,
\end{equation}
implying that the linear average yields a different localization length
compared to (\ref{chir1}).

By repeating the same calculations for the expression (\ref{fs}), one
obtains that $f_s \simeq \exp (|n-n_0|/l_\infty)$, which is found to be in
agreement with the numerical results for the KR model as it implies $l_s = 2
l_\infty$.

However, relation (\ref{fs}) implies that the average $<|\phi_m(n)|^2>$
should be used rather than $\langle |\phi_m(n)| \rangle$. In such a case,
the result is 
\begin{equation}  \label{normcor}
\left\langle \left| \varphi_m (n)\right|^2 \right\rangle \, \sim \exp\left(
- \frac{\left| n-m\right| }{2l_\infty }\right) \quad ,
\end{equation}
which implies that $f_s\simeq \exp (|n-n_0|/2l_\infty)$. The correctness of
this expression is confirmed by the relation $l_s = 4 l_\infty$ that it
implies.

\subsection{\bf Numerical data}

In order to determine the asymptotic shape of the wave packet, we have
followed the evolution of an initially $\delta$-like packet for times $t\geq
120t_D$. The distribution $f_s(n)$ has been then obtained by averaging over
more than 150 realizations for several $b$-values in the range $b= 4 \div 12$%
. The results are reported in Fig.~8 with the by-now-standard scaling
hypothesis, 
\begin{equation}
{\tilde{f}_s(x)}=b^2f_s(n),\,\,\,\,x=n/b^2 \quad ,  \label{scpro}
\end{equation}
that is once more confirmed by the good data collapse.

A peculiarity of all our simulations is that $f_s(n_0)$ is larger than the
neighboring values by approximately a factor 3. The reason of this apparent
anomaly can be traced back to the specific $\delta$-like shape of the
initial packet (that implies Eq.~(\ref{fs})) and to the spatial random
structure of the eigenvectors. Indeed, the latter assumption, together with
the observation that only a finite number $L$ of channels effectively
contribute to the sum in Eq.~(\ref{fs}), leads to 
\begin{equation}  \label{aver}
f_s(n_0)\approx L\,\left\langle \varphi ^4(n_0)\right\rangle
\,\,;\,\,\,\,\,\,\,f_s(n\neq n_0)\approx L^2\,\left\langle \varphi
^2(n)\right\rangle \left\langle \varphi ^2(n)\right\rangle \quad .
\end{equation}
Since it is known that $\left\langle \varphi ^4\right\rangle =3/L$, while $%
\left\langle \varphi ^2\right\rangle =1/L$ for the eigenfunctions of
matrices belonging to the Gaussian Orthogonal Ensemble \cite{Brody}, we
obtain that $f_s(n_0)/f_s(n) = 3$ for $n$ close to but different from $n_0$.
This value of the ratio is in pretty good agreement with the above mentioned
numerical estimate.

Moreover, the numerical results reported in Fig.~8 strongly suggest that the
decay of $f_s(n)$ in the vicinity of $n_0$ is definitely faster than in the
tails. Therefore, it is very tempting to compare this data with the
theoretical dependence derived for 1D disordered models (Eq.~(\ref{gogolin}%
)). The best fit of the only free parameter gives $l_m \sim 0.29 $; the
corresponding curve is shown in Fig.~8 (see the solid line). The very good
agreement between the numerical results and the analytical curve over a
broad range of $x$-values suggests that a properly modified theory to
include the determination of the mean free path from first principles should
be able to account for the asymptotic properties of packets in quasi-1D
systems as well.

One should also notice that the dependence of the slope on the distance from
the center $n_0$ of the packet is an entirely new feature with respect to
the analogous problem in the KR \cite{CS86,CIS88}, where no evidence of two
distinct regions of localization has been found. The reason of this
discrepancy is not clear: on the one hand it is possible that this reflects
an actual difference between the two models, on the other hand, it is
possible that the accuracy of the numerical data for the KR \cite{CS86} is
not enough to reveal this peculiarity in the associated profile $f_s(n)$.

A further difference is the variation of the localization length of the
eigenstates with the energy in BRM (see Eq.~(\ref{ll})). Far from the center
of the packet, we expect that the decay is dominated by the longest
localization length $l_\infty (0)=2/3$ (in $b^2$ units). On the other hand,
from Eq.~(\ref{asymp}), we find that $l_s=8l_m\approx 2.32$, which is only
slightly smaller than $4l_\infty (0)\approx 2.66$. Accordingly, the equality 
$l_s=4l_\infty $ found in the phenomenological theory for KR and explained
by invoking the presence of strong fluctuations of the individual
eigenstates \cite{CS86} appears to hold also in the present case. The small
deviation is presumably to be attributed to the not yet vanishing
contributions of more localized eigenstates. Instead, if we average over all
energies, we obtain $\langle l_\infty (E)\rangle =0.5$ (in $b^2$ units)
which results to a localization length $l_s=4\langle l_\infty (E)\rangle =2$
for the total wavefunction. Moreover, it is interesting to notice that a
direct determination of $l_s$ by fitting the profiles reported in Fig.~8
with a pure exponential law, yields $l_s\approx 2$: this means that the
multiplicative correction $1/|x|^{3/2}$ in expression (\ref{asymp}) is
essential for a correct estimate of $l_s$, if the range of $n/b^2$ is not
large enough.

A further more direct confirmation of the presence of strong fluctuations in
the various eigenstates is obtained by determining the localization length $%
l_s^{(l)}$ of the asymptotic packet from the logarithmic average $%
(\left\langle \ln (f_s)\right\rangle )$ of the single packets. The resulting
profile, reported in Fig.~9, yields $l_s^{(l)} \approx 1.3$ to be compared
with the value $l_s \approx 2.32$, obtained from the arithmetic average.
Interestingly enough, the ratio between the two lengths $l_s^{(l)}$ and $l_s$
for the shape of saturated wavepacket, is approximately equal to 2 as known
for the single eigenstates (see the previous subsection).

\subsection{\bf Fluctuations of the steady-state}

In this section we investigate directly the nature of wave-packet
fluctuations in the asymptotic regime. This enables us to test the
correctness of a conjecture relative to the single eigenvectors raised in
the context of the KR. Indeed, fluctuations are very important in that they
allow explaining the difference between the decay rate of the wave packet
and that of the eigenvectors.

Having in mind Eq.~(\ref{fluct}), we computed the logarithm of $f_s(n)$ and
studied its variance at a distance $\Delta n=6b^2$ from the center of the
packet (averaging also over a small window of 5 neighboring sites, under the
assumption that the fluctuations are nearly constant in such interval). As a
result, we have found that the distribution function of $y\equiv \ln (f_s)$
is, with a good accuracy, a Gaussian (see for instance the histogram
reported in Fig.~10 which refers to the case $b=5$ and is the result of
5,000 simulations with independent realizations of the disorder). This
represents a first confirmation that the hypotheses made in the KR can be
profitably carried over to the present model.

A more complete information is obtained by studying the fluctuations of $y$
for different values of $\Delta n$. More precisely, we have computed the
variance 
\begin{equation}  \label{var}
\sigma ^2(n)=\langle (\ln f_s(n))^2 \rangle - \langle \ln f_s(n) \rangle ^2
\end{equation}
in the steady-state regime. The results for the cases $b=5\div 8$ are reported
in Fig.~11 with the scaling assumption $\sigma ^2(n)=b^2 \sigma ^2(x)$,
where $x\equiv \Delta n/b^2$. The data shows that for large $x$, the
variance grows linearly with $x$, indicating that the logarithm of the
profile diffuses around the average value. This is a further confirmation of
the validity of a relation of the type (\ref{fluct}) for the wave packet and
it is strengthened by the observation that the estimated slope is
approximately equal to 1 in scaled units.

This behaviour is also analogous to what happens in 1D disordered models,
where it is assumed that the logarithm of the absolute value of the Green
function $\ln |G(m,n;E)|$ has a Gaussian probability distribution with the
mean equal to $-|m-n|/l_\infty $ and the variance equal to $|m-n|/l_\infty$ 
\cite{E83}.

Another interesting observation concerns the central part of the packet
(i.e., $|x|\leq 1 $), where the variance $\sigma ^2$ is almost constant,
indicating that the amplitude values are essentially independent of one
another (see also the inset in Fig.~11). Accordingly, all numerical findings
do confirm the conjectures that have been so far utilized to present all the
features of wavepacket diffusion in a coherent manner. Unfortunately, so far
there are no analytical results concerning the structure of the eigenstates
in the middle of their localization region, even in the well-studied 1D
Anderson case.

\section{\bf Conclusions}

In the present paper we have studied the evolution properties of wavepackets
in quasi 1D disordered media described by tight-binding Hamiltonians with
long-range random interactions. We have found that the wavepacket: (i) first
spreads ballistically, over a time scale of the order $t\sim 1/b^{0.5}$,
which becomes negligible in the limit $b \to \infty$; (ii) exhibits a
diffusive behaviour, for times of the order $t\sim b^2$; (iii) finally, for
times larger than $t_D\geq b^2$, stops spreading remaining asymptotically
localized.

The scaling properties of the spread of the packet are different in the
ballistic (see Eq.~(\ref{tbscale})) and diffusive (see Eq.~(\ref{tdiff}))
regime. Beyond the ballistic regime, we propose the heuristic formula (\ref
{fit}) to effectively describe both the diffusive spreading and the eventual
saturation. The most interesting feature of Eq.~(\ref{fit}) is the
prediction of a power-law convergence of the wavepacket width $M$ to its
asymptotic value, the deviation going to zero as $1/t^\beta $ with $\beta
\simeq 0.8$. The parameter $\beta$ accounts for the repulsion between those
eigenstates which are effectively excited by the initial wavepacket. More
precisely, the slow convergence of $M$ is attributed to effect of Mott
states, that are expected to give eventually rise to a logarithmic time
dependence, that cannot be numerically observed.

The ``short'' time diffusive regime has been investigated by computing $%
dM/dt $ when a small amount of time-dependent disorder was superimposed to
the quenched disorder. As a result we found that the diffusion coefficient
in dimensionless units is approximately four times larger than the
localization length, after averaging over the energy dependence. The
presence of noise with strength $g^2$ of the order of a critical value $%
g_{cr}^2\sim 1/b$, destroys quantum coherence and recovers classical
diffusion.

Another issue addressed in this paper concerns the fluctuations of the size
of the packet over different time scales. We found that for both diffusive
and saturation time scales, the relative amplitude of the fluctuations
scales as $\mu\equiv \Delta M/M \sim 1/b^{\eta}$ where $\eta \simeq 0.7$. We
confirmed this law by taking the Fourier power spectrum $|U(\omega)|^2$ of
sample to sample fluctuations of $M$. This result, besides confirming the
anomalous scaling behaviour provides evidence for the ergodicity of the
evolution.

The study of the Fourier power spectrum shows that the tails of $|U(\omega
)|^2$ have a Lorentzian-type form. At small frequencies, a weak singular
behaviour for $|U(\omega )|^2$ has been detected; this result is in
agreement with the power-law convergence of $M$ to its asymptotic value.
Furthermore, we have been able to extract the cross-over frequency $\omega_c$%
, where deviations from $1/\omega^2$ start in $|U(\omega)|^2$. Such a
frequency is two orders of magnitude smaller that the mean average
separation between the energy levels which participate to the evolution of
the wavepacket. This is a further confirmation of the participation of Mott
states to the long-time evolution.

For what concerns the asymptotic shape of the wavepackets, we found that the
scaling law (\ref{scpro}) is well verified already for relatively large $b$
values. Moreover, the analytic expression derived in the context of the 1D
Anderson model (see Eq. (\ref{gogolin})) reproduces pretty well the shape of
the average profile, upon fitting a single parameter. A further interesting
result of our numerical analysis concerns the difference by a factor 4
between the decay rate of the asymptotic profile around the origin and that
along the tails. Moreover, the difference between the localization length
estimated from the average of the logarithm of the profile and that obtained
from the linear average of the profile confirms the relevance of
fluctuations. This observation is reinforced by the asymptotic linear growth
of the variance $\sigma ^2$ of the logarithm of the profile versus the
distance from the center of the packet.

\section{\bf Acknowledgments}

We acknowledge useful discussions with B. Chirikov, Y. Fyodorov, A. Mirlin,
S. Ruffo and D. Shepelyansky. (F.M.I) is grateful to A. Bulgac for
illuminating discussions of the related problems in nuclear physics
application. (T.\thinspace K.) and (G.\thinspace P. \thinspace T.)
acknowledge discussions on the Anderson localization with C. Soukoulis and
E. N. Economou. They also acknowledge partial support from Human Capital and
Mobility grant CHRXCT930331. (T.\thinspace K.) acknowledge the support of
Grant CHRX-CT93-0107 and also wishes to thank Instituto Nazionale di Ottica
for the kind hospitality during the 1995 fall. Partial support by grant
INTAS-94-2058 is also acknowledge by (F.M.I).

\newpage
\centerline{\bf Figure Captions} \vspace{1.0cm}

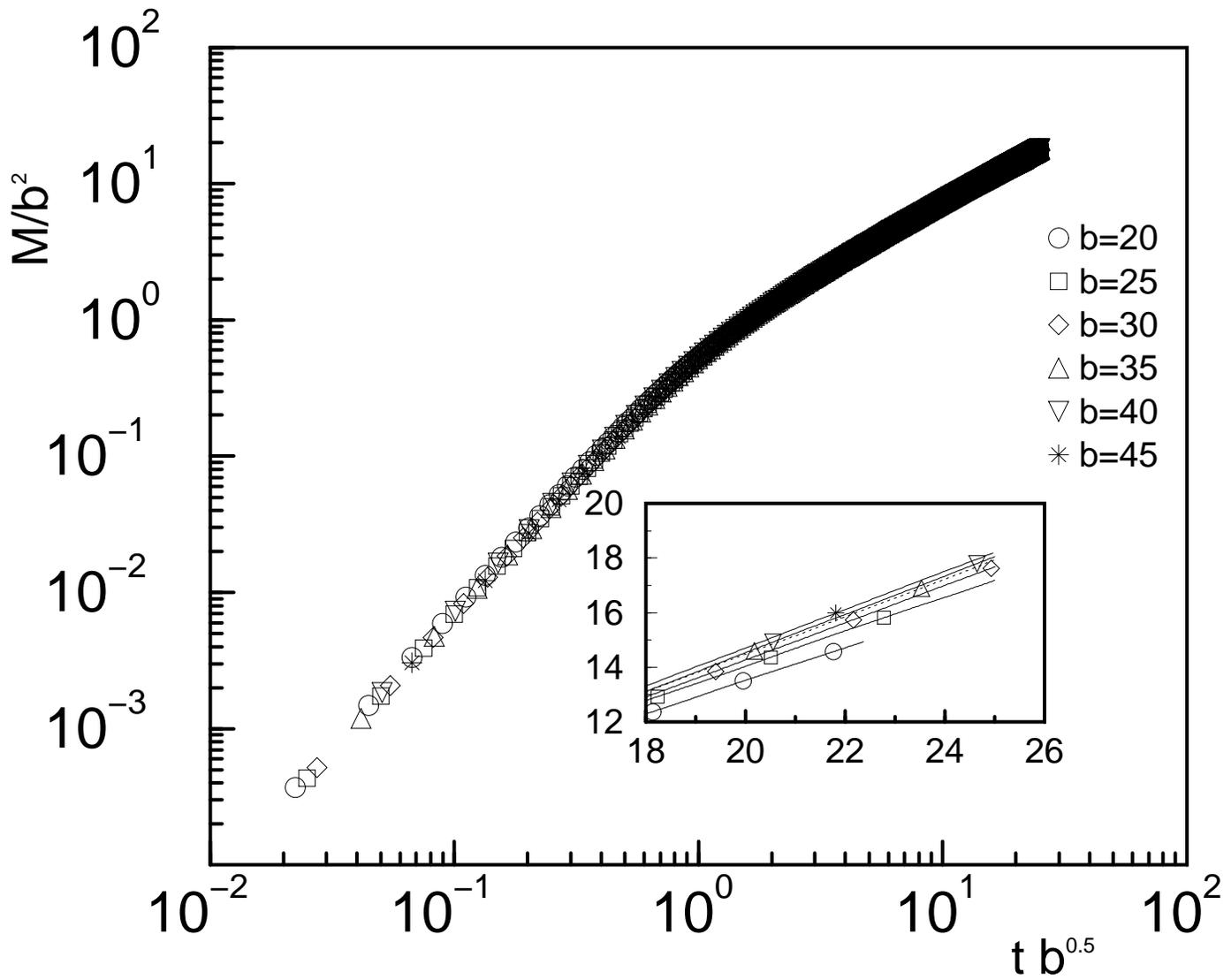
{\bf Fig. 1} Scaling of $M(b,t)$ vs. time for the ballistic time scale. The
reported values of $b$ are $b=20,25,30,35,40,45$. The inset shows the lost
of scaling for $t>t_b$.

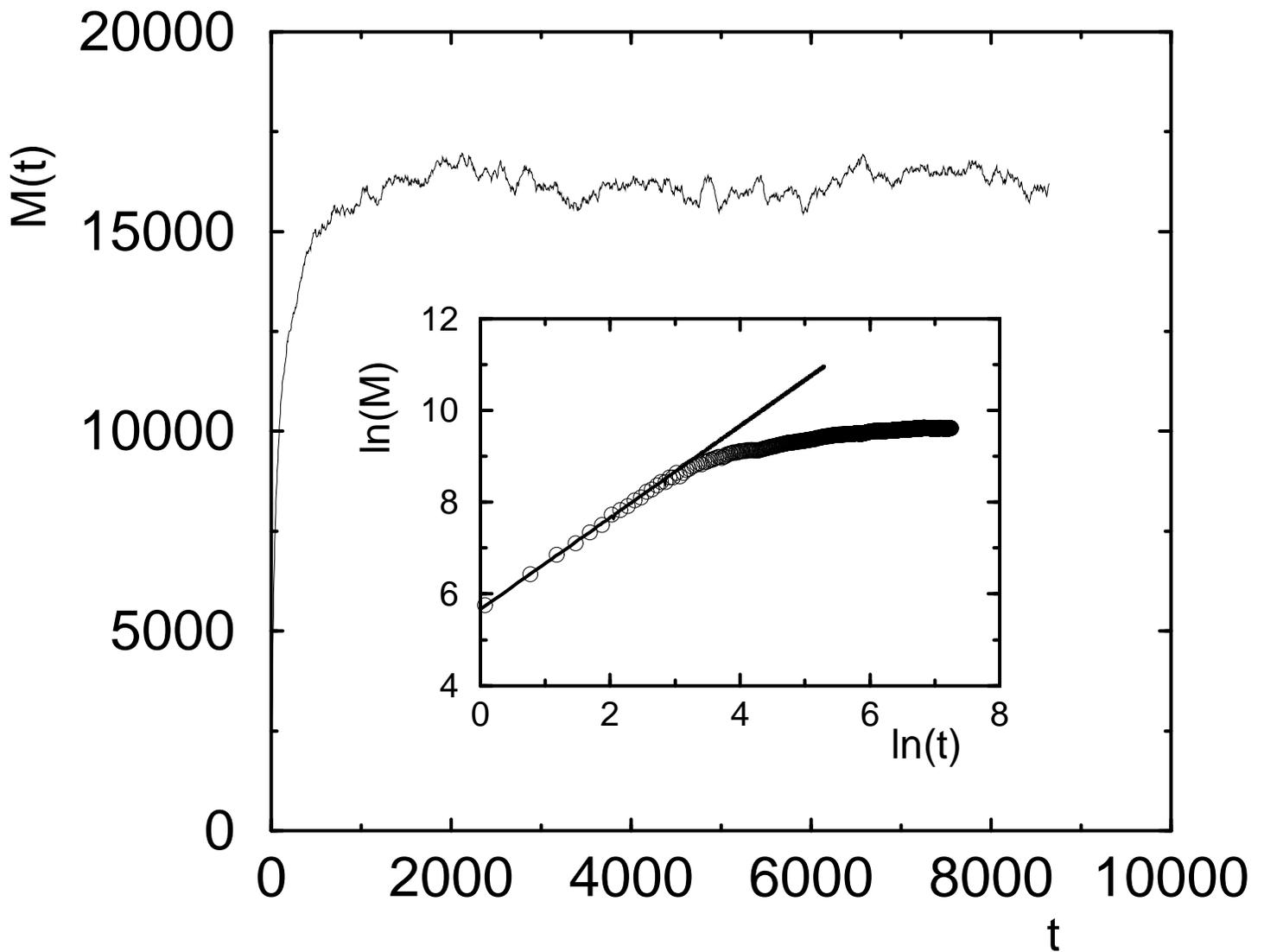
{\bf Fig. 2} An example of wave-packet diffusion beyond the ballistic time
scale for $b=12$. The same curve is reported (in a doubly logarithmic plot)
in the inset, magnifying the early stages of diffusion, to testify the
linear behaviour.

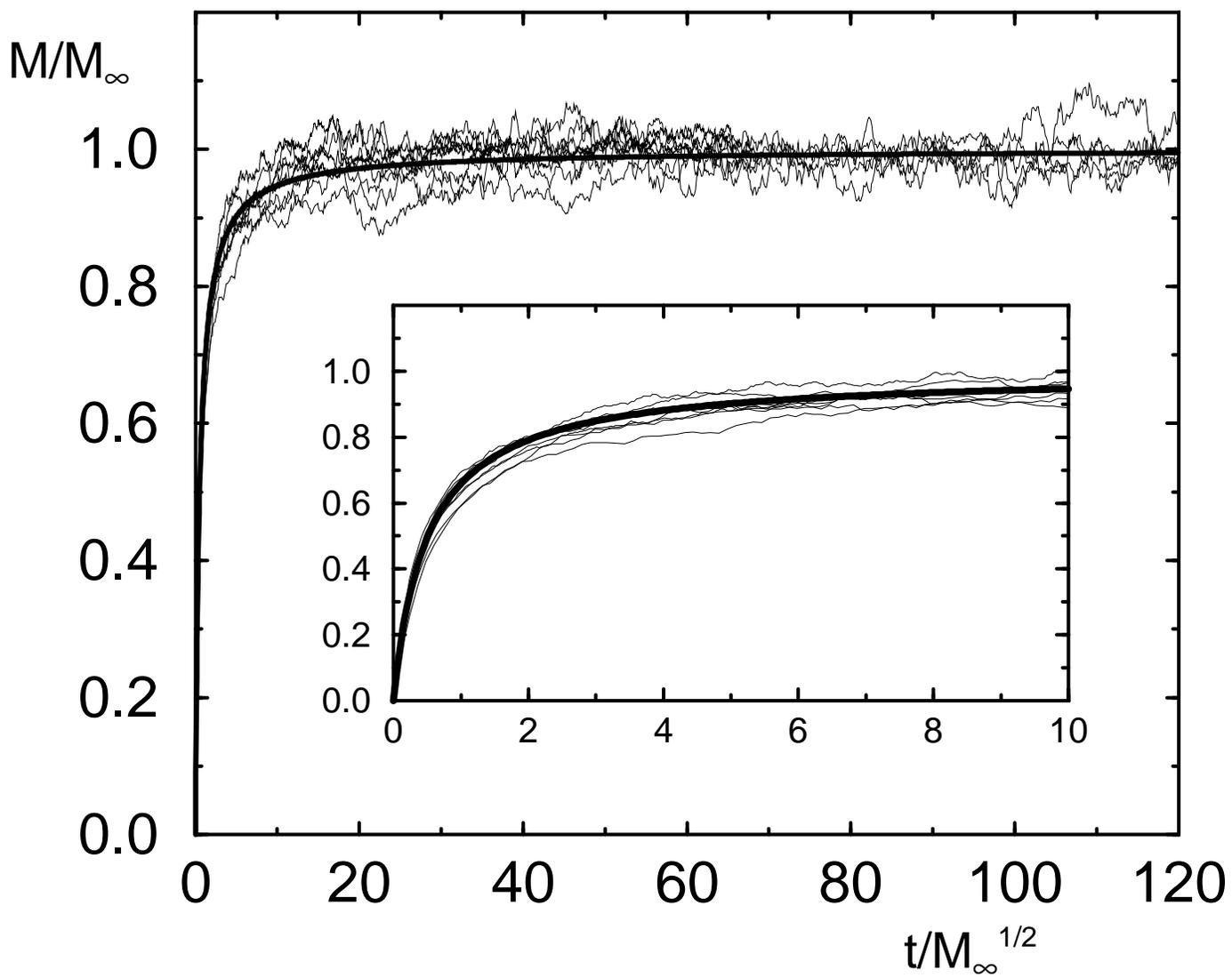
{\bf Fig. 3} Mean square displacement $M$ for $b=4,5,6,7,8,10,12$. The
smooth curve corresponds to the phenomenological expression (16). In the
inset the same quantities are shown for shorter times.

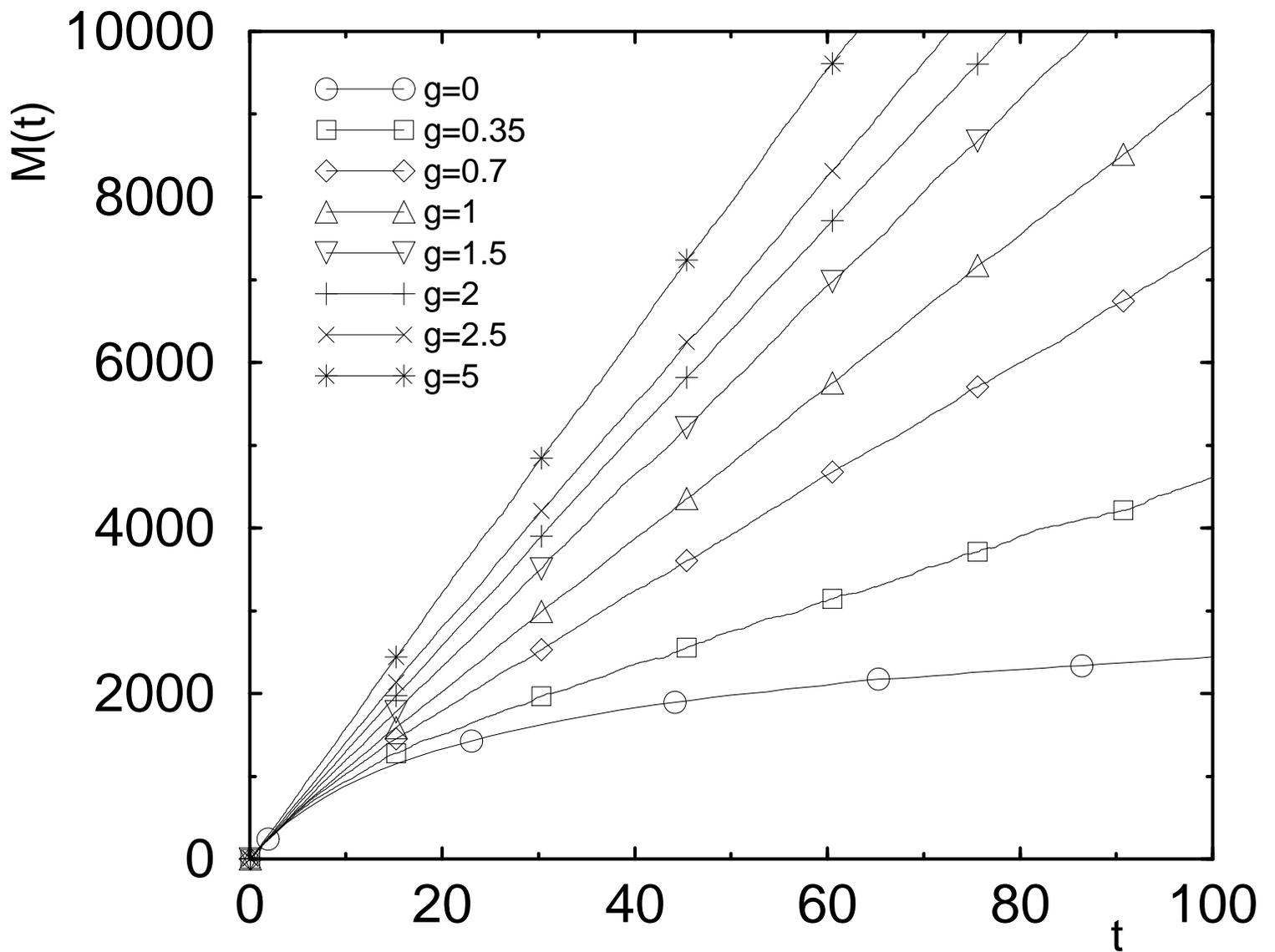
{\bf Fig. 4} Diffusion of packets with additional noise for $b=8$. The
diffusive constant is determined from simulations performed for different
noise strengths $g^2$.

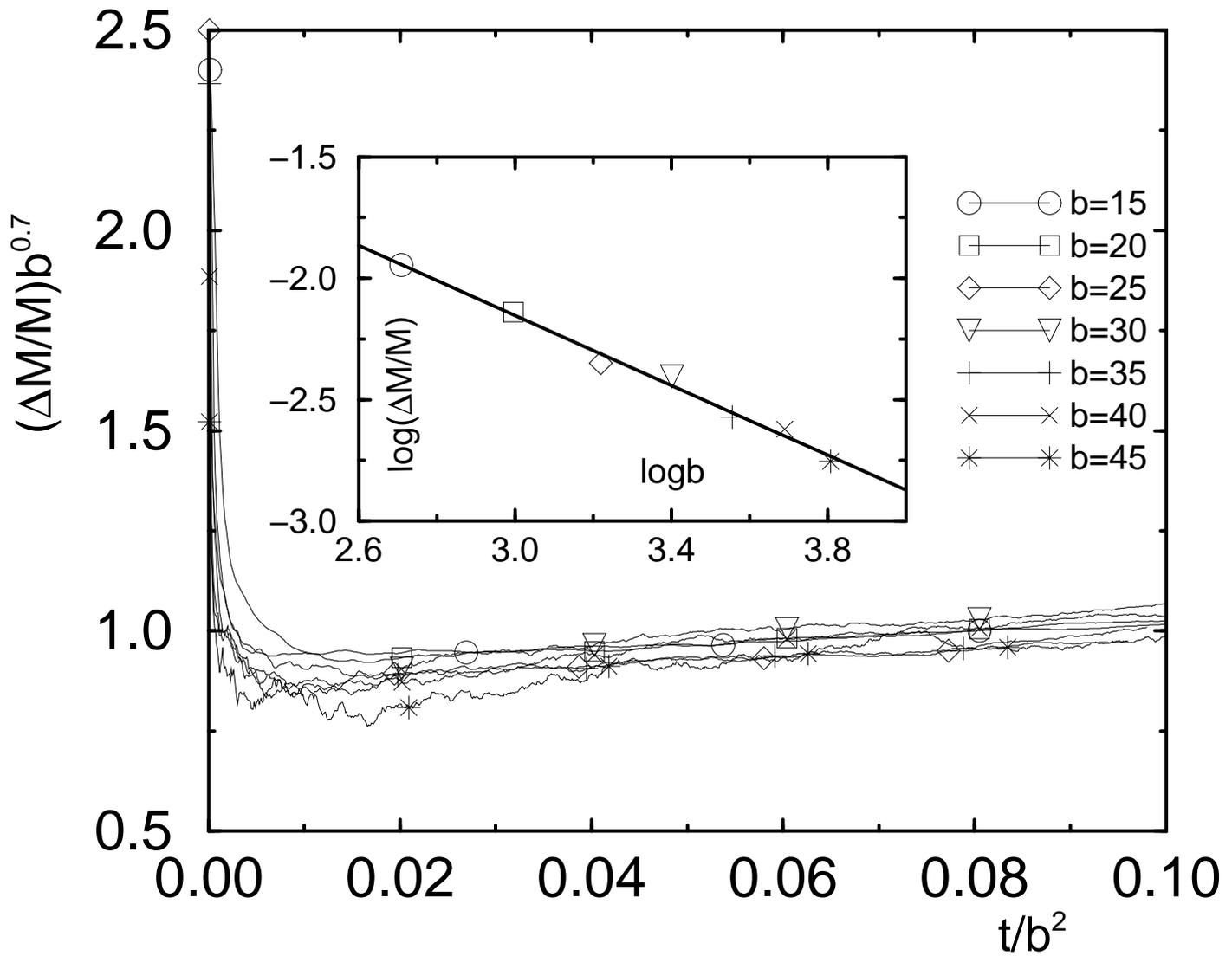
{\bf Fig. 5} Relative fluctuations on the diffusive time scale. The data is
scaled with respect to the law (20). In the inset, a least square fit for
the same quantity and different $b$-values is shown; the time is fixed, $%
t=0.1b^2$.

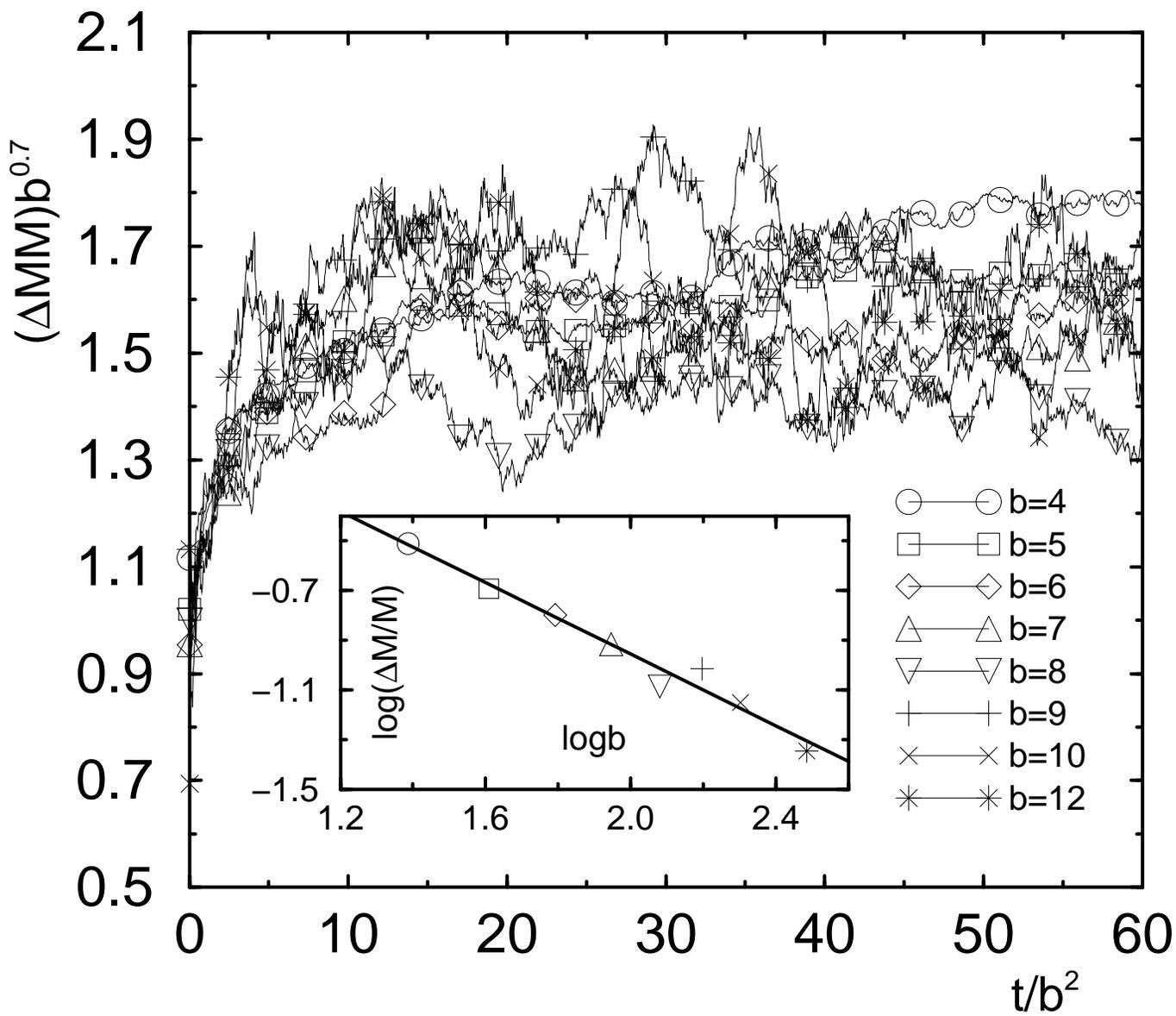
{\bf Fig. 6} Relative fluctuations in the saturation regime. The values of $%
b $ are $4,5,6,7,8,10,12$. In the inset, a least square fitting is shown for
the same quantity which now is averaged over time from $t=20b^2$ up to $%
t=120b^2$.

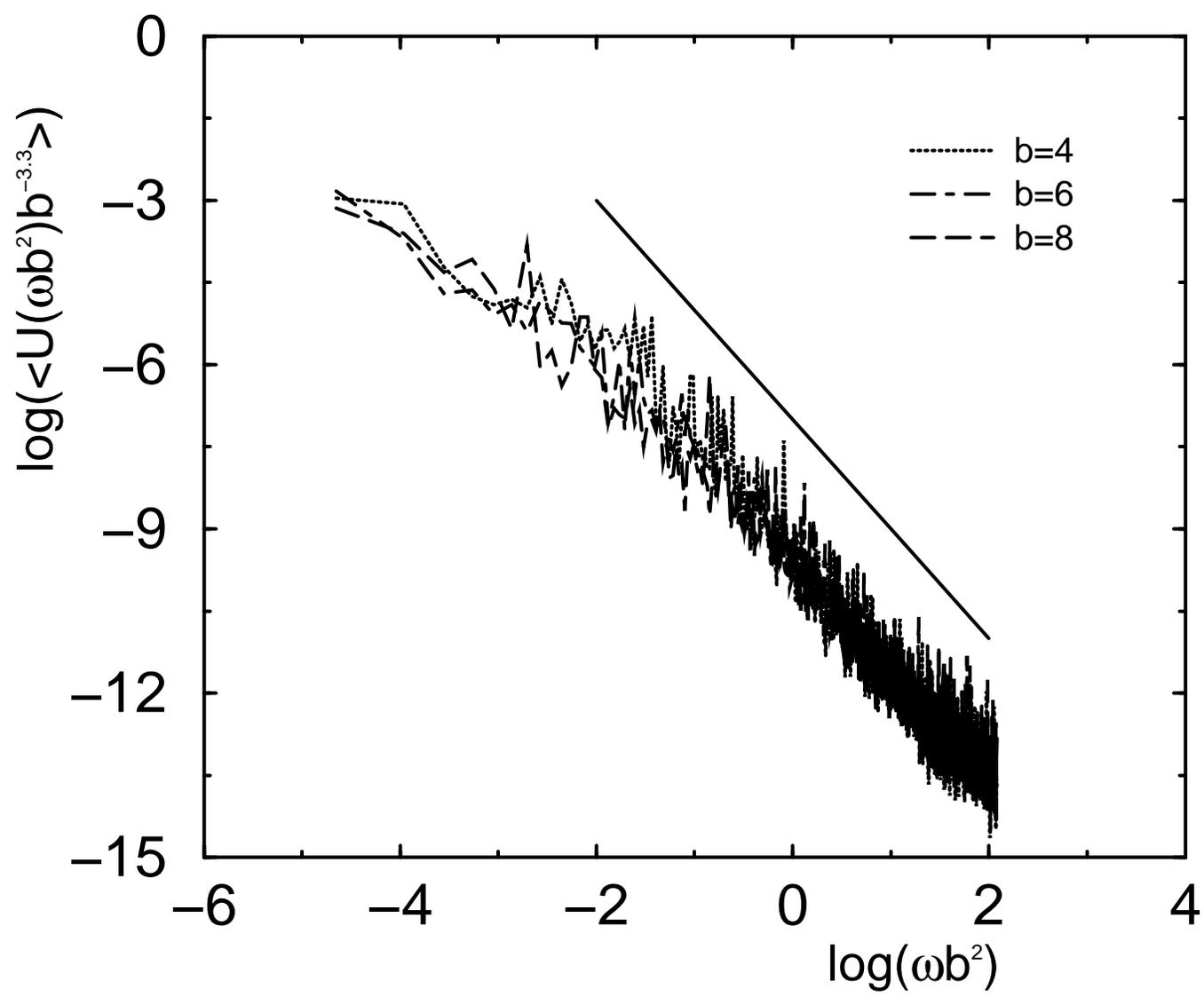
{\bf Fig. 7} The Fourier spectrum $\langle |U(\omega)|^2 \rangle$ for the
cases $b=4,6,8$ and for times $t> 20b^2$.

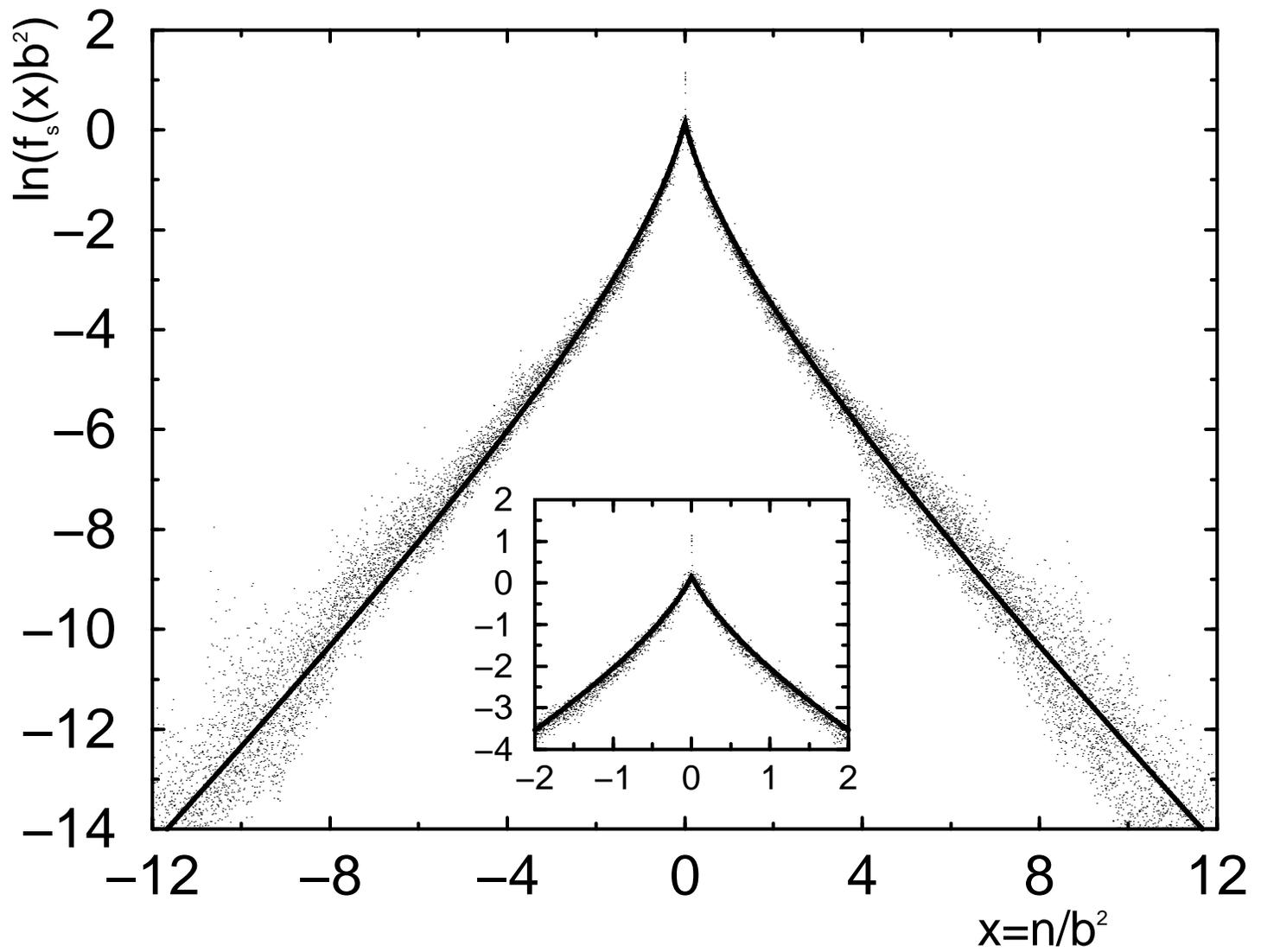
{\bf Fig. 8} Asymptotic average profile of the wavepacket for $b=4 \div 12$
after rescaling. The inset shows the behavior near to the maximum; smooth
lines follow from the theoretical expression (\ref{gogolin}).

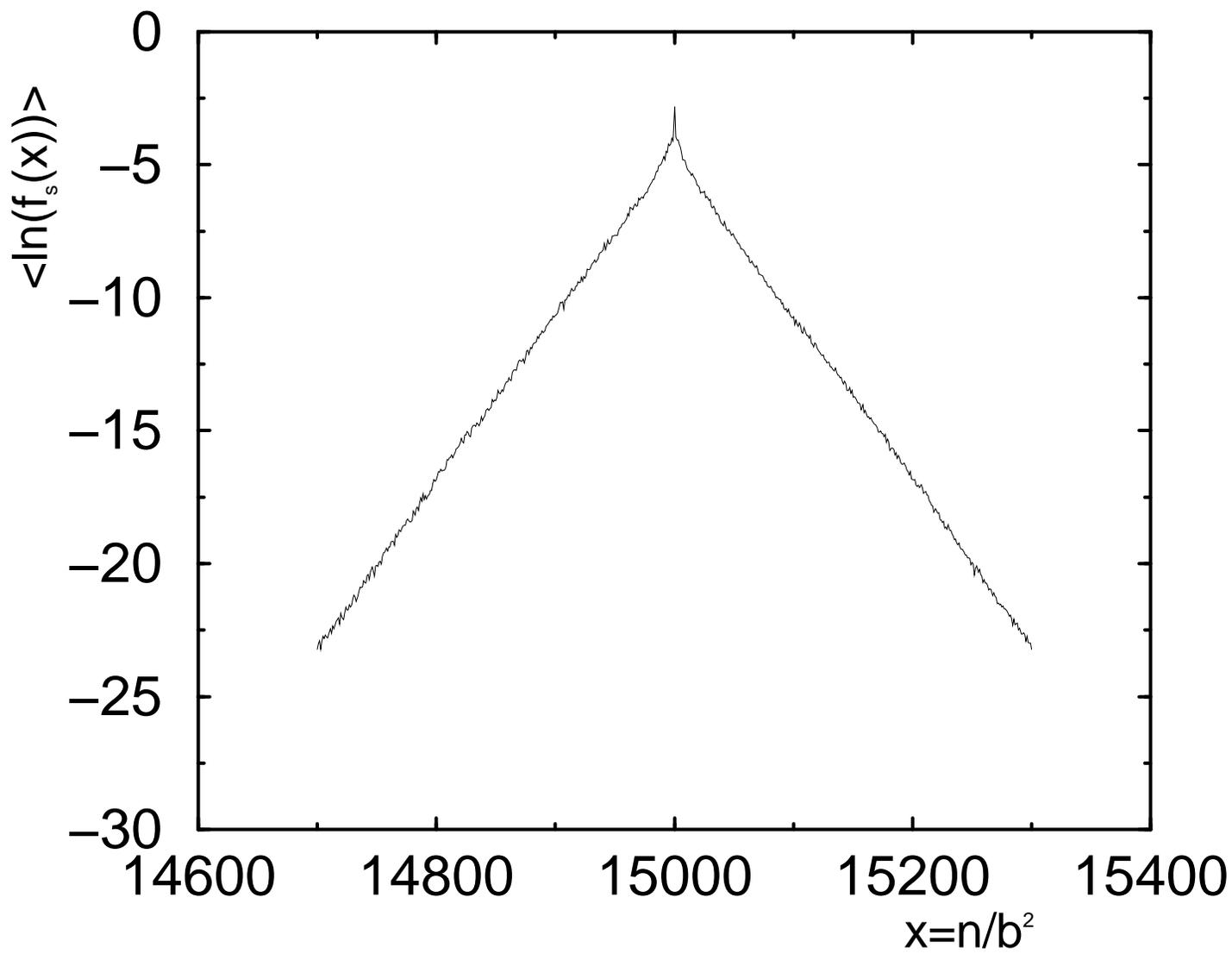
{\bf Fig. 9} Logarithm of the steady-state $f_n$ for $b=5$.

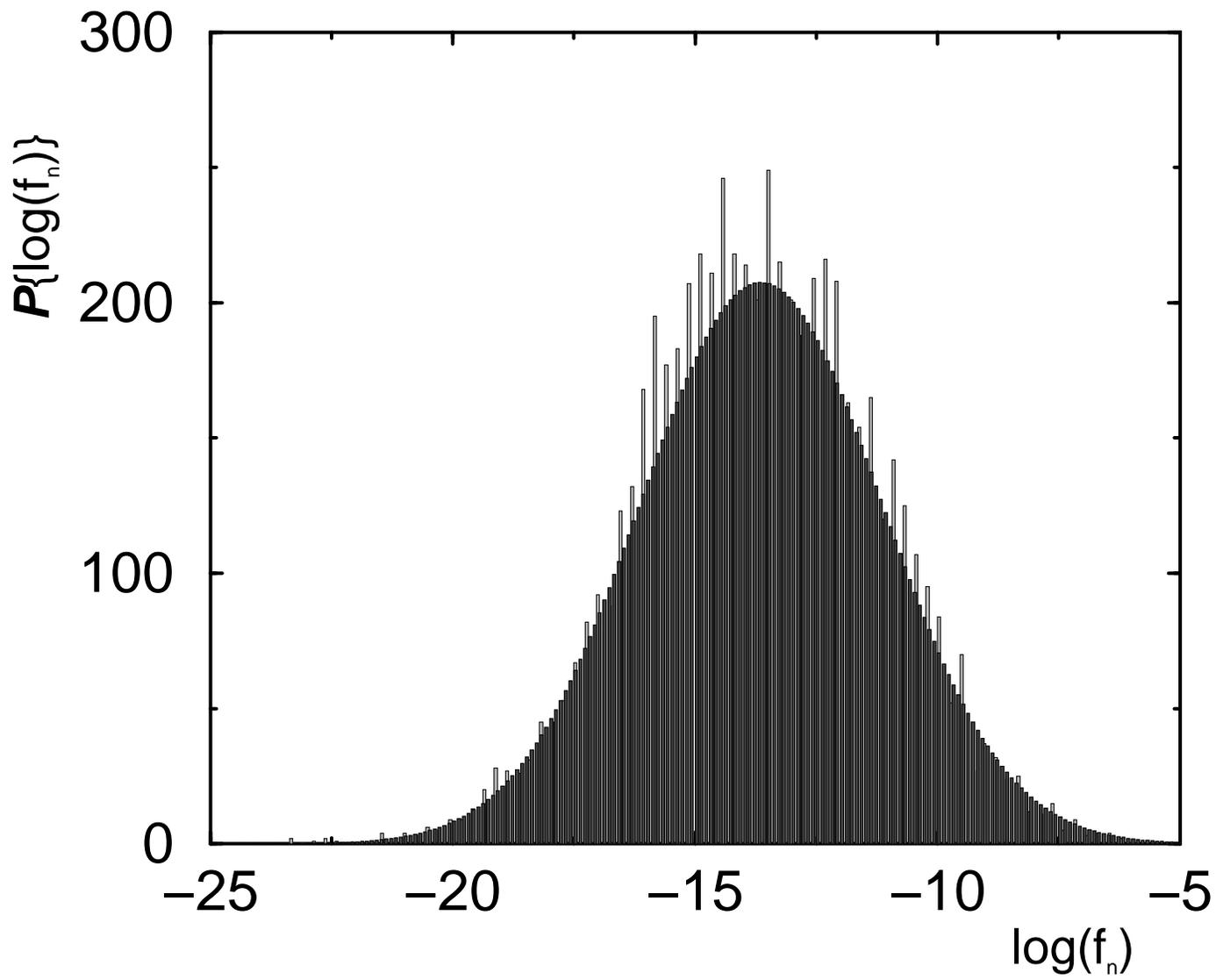
{\bf Fig. 10} Distribution of $\log f_n$ for $b=5$.

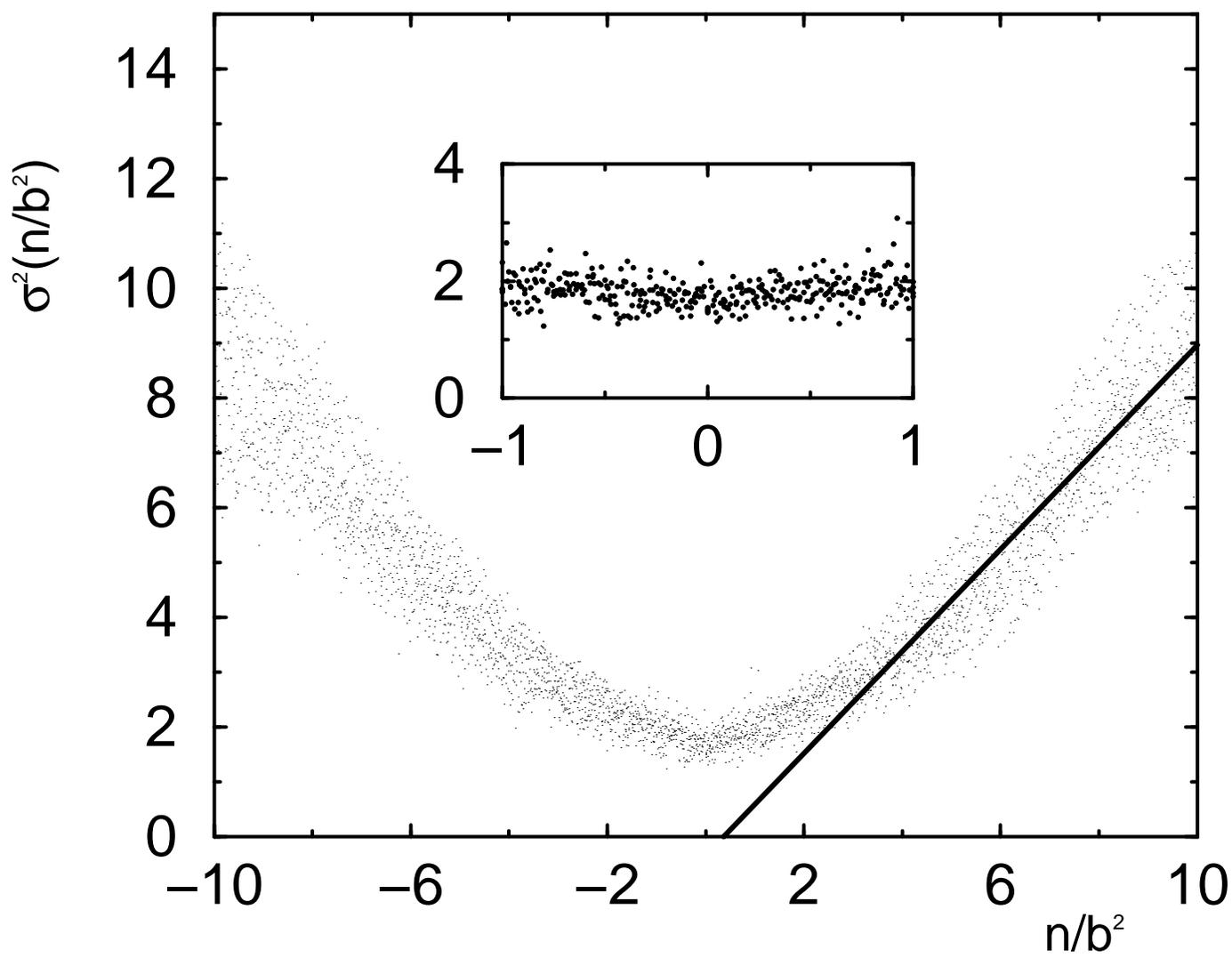
{\bf Fig. 11} Variance of $\log f_n$ in the steady-state after rescaling the 
$x$ axis for $b=5,6,7,8,10$. The straight line is the fit for large values
of $x$. The inset shows the behavior close to the maximum.


\begin{references}
\bibitem{note}  The only condition is that the amplitude of the matrix
elements has to decrease sufficiently fast in such a way that the second
moment of the envelope is finite.
\end{references}
\end{document}